# SELECTIVITY CORRECTION IN DISCRETE-CONTINUOUS MODELS FOR THE WILLINGNESS TO WORK AS CROWD-SHIPPERS AND TRAVEL TIME TOLERANCE

**Tho V. Le**
Ph.D. Student
Lyles School of Civil Engineering
Purdue University
550 Stadium Mall Drive, West Lafayette, IN 47907, USA
Tel: +1 765 586 2836; Email: le39@purdue.edu

**Satish V. Ukkusuri, Corresponding Author**
Professor
Lyles School of Civil Engineering
Purdue University
550 Stadium Mall Drive, West Lafayette, IN 47907, USA
Tel: +1 765 494-2296; Fax: +1 765 494 0395; Email: sukkusur@purdue.edu

Word count:  5,606 words text + 6 tables/figures x 250 words = 7,106 words

Submission Date August 1st, 2017

## ABSTRACT

The objective of this study is to understand the different behavioral considerations that govern the choice of people to engage in a crowd-shipping market.  Using novel data collected by the researchers in the US, we develop discrete-continuous models. A binary logit model has been used to estimate crowd-shippers' willingness to work, and an ordinary least-square regression model has been employed to calculate crowd-shippers' maximum tolerance for shipping and delivery times. A selectivity-bias term has been included in the model to correct for the conditional relationships of the crowd-shipper's willingness to work and their maximum travel time tolerance. The results show socio-demographic characteristics (e.g. age, gender, race, income, and education level), transporting freight experience, and number of social media usages significant influence the decision to participate in the crowd-shipping market. In addition, crowd-shippers pay expectations were found to be reasonable and concurrent with the literature on value-of-time. Findings from this research are helpful for crowd-shipping companies to identify and attract potential shippers. In addition, an understanding of crowd-shippers - their behaviors, perceptions, demographics, pay expectations, and in which contexts they are willing to divert from their route - are valuable to the development of business strategies such as matching criteria and compensation schemes for driver-partners.

*Keywords*: Crowd-shipping, Willingness to work, Last-mile delivery, On-demand delivery,
              Selectivity correction, Discrete-continuous model

## INTRODUCTION

E-commerce has seen a remarkable increase in the last decade, and is projected to grow significantly in the future (*1*). The number of online shopping orders per capita in 2013 was 28.7, 23.1, 18.2, and 13.6 in the UK, China, Germany, and the US, respectively (*2*). Among online product categories, electronics, fashion, apparel, and books are reported as the most commonly purchased items (*3*). A large variety of retailers - supermarkets, foods, beverages, apparel stores, bookstores, stationery stores, drug stores, electrical supply stores, florist shops, and souvenir shops - deliver small packages via regular transporters, carriers, or retail staff. As a result, logistics carriers have delivered a considerable number of small packages. Remarkably, the express delivery service for e-commerce purchases in China had a massive rise of 820% from 2009 to 2014 (*4*). In the report for Barclays (*5*), Conlumino revealed that letterbox-sized packages and small parcels were around 60% of all UK online purchased deliveries in 2013. Letterbox-sized packages can fit through a standard UK letterbox, while small parcels are no larger than a standard UK shoebox. Moreover, Bringg stated that there were over 2 billion local deliveries in the US alone in 2016 (*6*). As such, a huge demand for the delivery of small packages in urban areas can be observed.

In response to the demand for delivery services, the rise in e-commerce, and improved internet connection and smartphone technologies, a substantial number of crowd-shipping firms has been established worldwide. One challenge for crowd-shipping companies is to provide sufficient supplies to accommodate this demand. Therefore, identifying the potential supply from a third-party is crucial for crowd-shipping companies. One possible solution is to attract people who travel anyway to utilize the unused capacity in their vehicles to transport freight. In fact, there is a potential to make use of available vehicle volume for delivery freight. The US National Household Travel Survey from 2009 revealed the national average vehicle occupancy for all trip purposes was 1.67 (person miles per vehicle mile) (*7*). In addition, data from some other US cities also confirmed similar statistics (*8-10*), as can be seen in Table 1.

**Table 1.** Average vehicle occupancy

| National/City | Year | Sample size of the survey | Average vehicle occupancy |
|---|---|---|---|
| USA | 2009 | 113,101 | 1.67 |
| Knoxville (TN) | 2008 | 11,522 | 1.70 |
| Chicago (IL) | 2007 | 10,552 | 1.67 |
| Atlanta (GA) | 2011 | 10,278 | 1.85 |

Moreover, delivery distance information is helpful for crowd-shipping companies' operational strategies (e.g. to alert potential crowdsourced drivers of jobs within their distance preferences). The maximum and minimum distances computed from a one-day data set of requests for freight delivery in Shanghai, China were 87.5 km and 0.18 km, respectively (*11*). The delivery requests were collected from traditional logistics carriers and local shops. The traditional logistics carriers outsourced e-commerce packages requested to be picked up from their local service branches, while the local shops requested online-to-off-line package deliveries. The average distance for all delivery trips was 6.5 km (i.e. 4 miles). The average delivery distance of packages ordered from local shops was only about 3 km (i.e. 1.86 miles).

Potential crowd-shipping supply, however, is not yet clear on various aspects. The following research questions remain: "Who is willing to work as a crowd-shipper? Are there any specific socio-demographic characteristics associated with prospective crowd-shippers? What factors drive them? Do they have any preference for the types or owners of shipments? What is the maximum tolerance for travel time (TTT) or distance that crowd-shippers would accept to divert from their typical routes for pickup and delivery purposes? What factors influence this decision? How much do crowd-shippers expect to be paid (ETP)?" (*12*). However, clear research based answers are not yet available. Accordingly, this research will investigate these questions with discrete-continuous approach models. These models examined prospective crowd-shippers' willingness to work (WTW) as well as the maximum TTT of respondents who are willing to work as crowd-shippers. This paper provides crowd-shipping companies a better understanding of effective and efficient system operations. This information will help crowd-shipping firms to recruit part-time drivers, understand when to work, how much to work, which circumstances affect prospective crowd-shippers' WTW, and set up operational strategies (e.g. matching and routing strategies, incentive for driver-partners).

This paper includes six sections. The introduction presents the background and motivations of this study. The literature review section illustrates its state-of-the-art methodology. Section 3 features information from the study's data sources. The findings and insights are discussed in the estimation results section. Finally, the study is summarized in the conclusion section.

**LITERATURE REVIEW**

Since there is no available data on crowd-shipping operational practices, comprehensive questionnaire sets were designed to address the aforementioned research gaps. The survey design, descriptive statistics and preliminary insights from the data are presented by the authors in another paper (*12*).

This research was motivated by the questions related to survey responses (i.e. the WTW and TTT choices). A logit model was employed to estimate for the discrete choices of WTW as crowd-shippers or not. An ordinary least-square regression model was employed to calculate crowd-shippers' maximum tolerance for shipping and delivery times. The selectivity-bias term was added to the regression model to correct for the correlations of discrete and continuous variables.

A variety of studies in the field of transportation have used discrete-continuous approaches. Bhat and his colleagues published a series of works using this technique. For example, Bhat (2005) developed a multiple discrete-continuous extreme value model to test time-use allocation decision. The results showed the significant influence of demographics and employment patterns on time-use patterns (*13*). The multiple discrete-continuous model was employed by Bhat and his colleagues in other studies as well (*14 - 15*).

Relationships between variables (e.g. discrete and continuous) have been found in various transportation data sets. Hamed and Mannering (1993) developed a new method to correct these correlations. The choice to go home from work and the travel time to home were modeled as discrete-continuous models to which the selectivity correction term was applied. Selectivity bias was presented in the model since the travel time data was only available from the subset of respondents who decided to go directly home from work. The authors employed a binary logit model to analyze the activity/home choice model and an ordinary least-square regression model for the travel time from work to home. The results showed that the selectivity correction term was statistically significant; therefore, correlations are present in the data set (*16*). Numerous works have addressed methodological issues related to this topic (*17-25*). Our goal in this paper is to apply this methodology to an innovative dataset in crowd-shipping and obtain various insights related to WTW and TTT choices.

Research on the supply side of the crowd-shipping system is limited. Paloheimo et al. (2016) studied a crowdsourcing pilot program for delivering library materials (i.e. books and library media) in Finland, and found an average detour of 2.2 km per delivery. In addition, of all reasons for respondents to participate in the trial, "try something new," "make life easier for me," "support public service," and "support the environment" were reported as the major motivations (*26*). In another study, 64% of respondents were willing to transport parcels and 44% were motivated by ecological interest, according to Briffaz and Darvey (2016) (*27*).

Regarding the value of potential crowd-shippers' WTW, few relevant publications have been found. In a recent study, Miller et al. (2017) estimated the value of WTW, which is defined as giving up time and making profits. The WTW value was found to be higher than the typical willingness to pay values presented in the literature. Moreover, socio-demographic and attitudinal variables were reported to have a significant influence on the WTW decision (*28*). The paper of Miller et at. (2017) is among the earliest research on the topic; however, selectivity bias was ignored in the modeling approach. Failure to correct for the selectivity bias leads to significant limitations on the insights and conclusions drawn from the estimated results (*29*). Therefore, the goals of this paper are to provide objective and consistent results as well as contribute to the emerging field of crowd-shipping research.

## METHODOLOGY

In this study, respondents were first asked whether they were willing to work as crowd-shippers (discrete variable). If so, they were asked what is the maximum TTT (continuous variable) that they would accept to pick up and deliver packages. Those decisions are interrelated; therefore, the discrete-continuous models are the best fit to analyze the data (*29*). In addition, the interconnected discrete-continuous data is generally considered as a problem of selectivity. The observed data (i.e. TTT) was the outcome of a selection process related to the non-random sample of data from observed discrete decisions (i.e. WTW as crowd-shippers). The relationships of the decisions are illustrated in **Figure 1**.

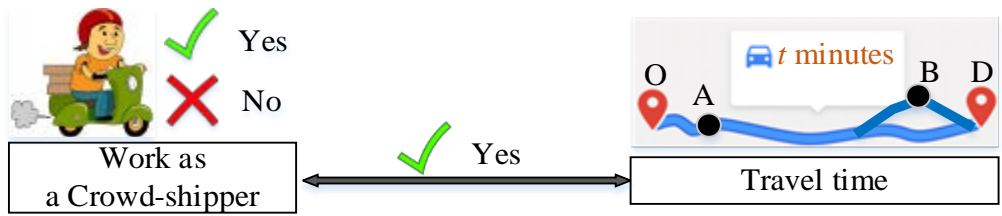

**Figure 1.** Relationships of the two decisions

The goal of this study is to identify which factors relate to the maximum TTT of respondents willing to work as crowd-shippers. The continuous TTT is defined as:

$$TTT_i = \alpha_0 + \alpha_1 X_i + \varepsilon_i \tag{1}$$

Where $TTT_i$ is the tolerance for travel time of crowd-shipper *i*; $\alpha's$ are estimable parameters; $X_i$ is a vector of respondent *i*'s social demographic variables; $\varepsilon_i$ is an unobserved term assumed to be normally-distributed. However, since this model is applied to the subset of respondents who are willing to work as crowd-shippers, $\varepsilon_i$ does not have a zero mean as assumed. Therefore, we need to use the selectivity bias to correct for the discrete-continuous models. The selectivity bias indicates a conditional TTT value given that respondents are willing to work as crowd-shippers. Several approaches have been developed to correct for such selectivity bias (*30*, *31*). Denote $E(\varepsilon_i \mid k)$ is a conditional mean of $\varepsilon_i$ given that respondent $i$ chooses to be a crowd-shipper (*31*).

$$E(\varepsilon_i \mid k) = \rho \left( \frac{\sigma_\varepsilon}{\sigma_\xi} \right) \Pi_{ik} \qquad (2)$$

Where $\sigma_\varepsilon$ is the standard deviation of the normally-distributed unobserved term $\varepsilon$; $\sigma_\xi$ is the standard deviation of the logistic unobserved term $\xi$ in the discrete choice model (Equation 6); $\rho$ is the correlation between $\varepsilon$ and $\xi$; and $\Pi_{ik}$ is defined as:

$$\Pi_{ik} = \left( \frac{(1-p_{ik}) * \log(1-p_{ik})}{p_{ik}} + \log(p_{ik}) \right) \qquad (3)$$

Where $p_{ik}$ is the probability of the decision of WTW as a crowd-shipper of a respondent $i$. Then

$$\varepsilon_i = E(\varepsilon_i \mid k) + \tilde{\varepsilon}_i \qquad (4)$$

Substituting (4) in in (*1*), the equation becomes:

$$E(TTT_i \mid k) = \alpha_0 + \alpha_1 X_i + E(\varepsilon_i \mid k) + \tilde{\varepsilon}_i = \alpha_0 + \alpha_1 X_i + \beta \Pi_{ik} + \tilde{\varepsilon}_i \qquad (5)$$

where $\beta$ is an estimated parameter which equals $\rho \left( \frac{\sigma_\varepsilon}{\sigma_\xi} \right)$; $\tilde{\varepsilon}_i$ has a conditional zero mean by construction.

In Equation (5), the parameter $\beta$ of the selectivity-bias term is estimated as a random parameter. As such, a parameter is estimated for each observation. The hypothesis under this assumption of a random parameter is the variety of behavioral observations. In other words, all observations used in this model are willing to work as crowd-shippers, and their TTT varies. This Equation (5) is then computed using the ordinary least-squares method.

The discrete-continuous model with the selectivity correction term is consequently solved in the following three steps:

1. Using a discrete-choice model to estimate a probability for each discrete decision (i.e. willingness to work as crowd-shippers). The data set from all respondents is employed in this step.
2. Using the outcomes from step one to estimate values of selectivity.
3. The regression model is employed to evaluate the continuous data. This model includes the computed selectivity variable from step 2 that corrects for the selectivity bias of the discrete-continuous decision process. Only a subset of data, from respondents who are willing to work as crowd-shippers, is used in this model.

The multinomial logit model is widely used in studies of choice modeling. One property of this model is an assumption of IIA, which is suitable for independent choices. Therefore, a multinomial logit model is commonly employed to infer the self-determined behavior of respondents. The utility of decision $k$ of a respondent is expressed as $U_k$.

$$U_k = V_k + \xi_k \tag{6}$$

Where $V_k$ is the observed utility, and $\xi \sim IID\,Gumbel\,(0,1)$. The choice model is then written as:

$$P_k\left(V_k^* > V_k\right) = \frac{\exp(V_k)}{\sum_{k=1}^{K}\exp(V_k)} \tag{7}$$

In this study, the multinomial logit model is collapsed to a binary logit model since there are only two alternatives (i.e., willing to work as crowd-shippers or not) in the choice set.

## DATA SOURCE

The data set used in this study was collected from a US survey spanning from February to April 2017. The survey was designed to understand the behavior of stakeholders (e.g. requesters and prospective crowd-shippers) and assumed the availability of crowd-shipping services in the logistics market. There were 1,176 responses, but the final data set only includes 549 respondents, as some responses were incomplete or inconsistent. In the survey, shipping experience, as well as preferences and stated preference questions on crowdsourced delivery were asked. Respondents reported their experience of transporting freight for someone else in the past, and then were asked whether they were willing to work as crowd-shippers in the future given a number of contexts. The logic conditions were applied to direct respondents to the follow-up questions depending on their responses of "yes" or "no". For example, the respondents who were willing to work as crowd-shippers were asked for the maximum TTT they were willing to divert for picking up and delivering a package. Aside from responses to the hypothetical questions, the data set also includes socio-demographic characteristics, such as age, gender, race, education level, etc. Personal socio-economic data - income, number of children, number of adults in his/her household, and accommodation ownership - are also provided in the data set. The results show 78% of respondents are willing to work as crowd-shippers. The TTT average and standard deviation are 23 and 18 minutes, respectively, for 20-minutes of travel on the original route. TTT distribution is displayed in Figure 2. Readers can refer to the details of the questionnaire design, survey implementation, and descriptive variables in Le and Ukkusuri (2018) (*12*). In Table 2, only the characteristics of variables used in this study are summarized.

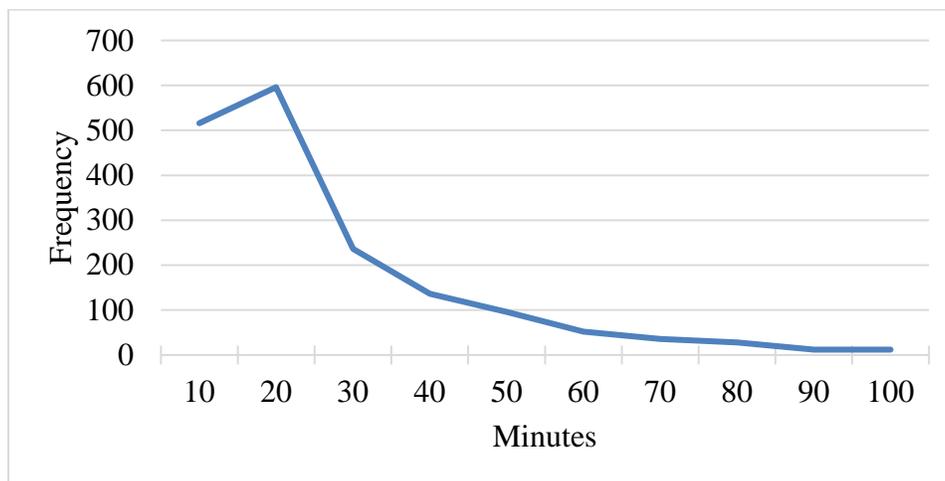

**Figure 2.** Distribution of tolerance for travel time (minutes)

**Table 2.** Descriptive statistics of explanatory variables

*percentages for indicator variables

| Variable Description | Unit | Min/ max or values | Mean (*Standard deviation*)* |
|---|---|---|---|
| **Numbers of observation (respondents): 2,196 (549)** | | | |
| Experience of transport freight for someone else. Dummy variable: 1- Yes; 0- No | NA | 0/1 | 25.70/74.30 |
| Will you work as a crowd-shipper? Dummy variable: 1- Yes; 0- No | NA | 0/1 | 78.32/21.68 |
| Age. Dummy variable: 1- If >30 years old; 0- Otherwise | NA | 0/1 | 65.26/34.74 |
| Male and number of children | NA | 0/5 | 0.29 (*0.75*) |
| African American/American Indian/Alaska native and income is less than $50,000/year | NA | 0/1 | 6.42/93.58 |
| Numbers of people in your household are >=65 years old | Number | 0/6 | 0.18 (*0.63*) |
| Having college degree or higher and income is less than $50,000/year | NA | 0/1 | 44.40/55.60 |
| Income | $1,000/year | 15/220 | 48.71 (*36.00*) |
| Household ownership. 1- Living in a house with mortgage; 0- Otherwise | NA | 0/1 | 20.00/80.00 |
| Total numbers of social media usages | Number | 0/10 | 4.00(*2.10*) |
| **Numbers of observations (respondents): 1,720 (430)** | | | |
| Maximum tolerance for travel time would you accept to pickup and delivery a package | Minutes | 1/100 | 23.40(*17.50*) |
| I can be a crowd-shipper during my commute | NA | 0/1 | 70.00/30.00 |
| ETP as a crowd-shipper | USD | 0/30 | 11.70 (*4.59*) |
| I can deliver whosoever packages or goods if I get paid | NA | 0/1 | 72.27/27.73 |
| Age. Dummy variable: 1- If <31 years old; 0- Otherwise | NA | 0/1 | 31.45/68.55 |
| Gender. Dummy variable: 1- Female; 0- Male | NA | 0/1 | 52.73/47.27 |
| African American/American Indian/Alaska native male | NA | 0/1 | 4.00/96.00 |
| Income is less than $30,000/year and deliver at weekday nights | NA | 0/1 | 26.80/73.20 |

The authors utilized NLOGIT 6 for all modeling work, including preliminary statistical analysis (as presented) and model building (*32*). The model development procedure and insights from the achieved results are provided in the following section.

## ESTIMATION RESULTS

The potential explanatory variables for the models were selected from theoretical and empirical studies on the sharing economy, ride-sharing and carpooling studies (*33-34*), and other crowd-shipping studies (*28*). In addition, hypothetical variables (e.g. transport freight during commute, transport freight for people who potential crowd-shippers know, ETP, and packages ownership) were also tested in the models during the model building process. It is noted that the correlations between variables were calculated to identify highly correlated variables and prevent multicollinearity issues before building the models. Pair-wise variables, including newly created variables and variables from the survey, were found to have no highly correlated; therefore, there is no issue of multicollinearity with the developed models. The results of the estimated models are presented in the sections that follow.

### Willingness to work as crowd-shippers model
As discussed, respondents selected whether or not they were willing to work as crowd-shippers (i.e. "Yes" and "No") from the choice set. Therefore, the binary-logit model was developed, and the WTW as crowd-shippers was selected as a dependent variable. Various explanatory variables were tested for statistical significance. There is no instrumental variable (i.e. endogenous variables associated with the corresponding alternative) that varies across alternatives. Explanatory variables only include respondents' socio-demographic characteristics. The results are presented in Table 3. All parameters (except the constant parameter) have plausible signs and a significance of more than 95%.

**Table 3.** Binary logit model estimation results of WTW as crowd-shippers and average marginal effects

| Variable Description | Coefficient | t-stat | Average marginal effect |
|---|---|---|---|
| Constant | 0.021 | 0.11 | - |
| Experience of transport freight for someone else. Dummy variable: 1- Yes; 0- No | 1.486 | 8.47 | 0.182 |
| Age. Dummy variable: 1- if >30 years old; 0- otherwise | 0.909 | 7.24 | 0.149 |
| Male and number of children | 0.320 | 3.12 | 0.049 |
| African American/American Indian/Alaska native and income is less than $50,000/year | 0.530 | 1.84 | 0.073 |
| Numbers of people in your household are >=65 years old | -0.207 | -2.48 | -0.032 |
| Having college degree or higher and income is less than $50,000/year | 0.583 | 4.10 | 0.088 |
| Income ($1,000/year) | -0.004 | -2.27 | -0.001 |
| Living in a house with mortgage | 0.432 | 2.80 | 0.063 |
| Total numbers of social media usages | 0.067 | 2.46 | 0.010 |
| Number of observation (respondents) | | | 2,196 (549) |
| Restricted Log Likelihood | | | -1148.010 |
| Log Likelihood at convergence | | | -1041.100 |
| Pseudo-R square | | | 0.093 |

*Note:* all variables are defined for the WTW as crowd-shippers.

Respondents who transported freight or goods for someone else in the past were willing to work in the crowd-shipping system. This may be due to the respondents' familiarity with the field and confidence to participate in a similar system. Moreover, the positive and statistically significant parameter of "age" suggests that people who are more than 30 years old tend to be crowd-shipping driver partners. Perhaps these respondents are more likely to have routine daily activities, therefore, they can more easily accommodate an additional task. The parameter of males who have multiple children is positive. This indicates that they are more likely to work as crowd-shippers, while females who have multiple children are less likely to do so. Males possibly consider themselves bread winners of the family and potentially have more flexible time schedules as compared to their female counterparts.

African American, American Indian, and Alaskan native respondents with a college degree or higher who earn less than $50,000/year are more likely to work for the crowd-shipping system. Earning about average or less than average income may motivate them to work as a crowd-shipper (e.g. the average income of the US in 2015 was $48,100 (*35*)). Low income people consider crowd-shipping as an additional opportunity to earn income. This is potentially an extra job with flexibility. In addition, the negatively significant income coefficient suggests that respondents who earn higher incomes are less likely to work as crowd-shippers, as expected.

Our findings also show that respondents who are living with elderly people are less likely to work as crowd-shippers. This is probably because of the constraints imposed by living with elderly family members. They may need to spend more time with and be available to the elderly citizens; therefore, it reduces the flexibility to participate in crowd-shipping. However, respondents who are living in mortgaged houses are more motivated to work as driver partners for crowdsourced delivery companies. This indicates the desire to potentially earn additional income to pay loans and other debts. Moreover, individuals who use more social media outlets are more likely to work as crowd-shippers. These people may be more technology savvy, familiar with using apps, and open to gigs in the sharing economy.

Different population group may be engaged in crowd-shipping by different reasons. Some may want "to earn money while looking for a full-time job". Some may be motivated to "maintaining steady income" or "earning more income" at a certain stage in their life. Others may work "to have more flexibility" or "to be your own boss" (*12*). Therefore, to promote crowd-shipping and address prospective driver partners, crowdsourced delivery companies could filter crowds by multiple criteria for their promotion and recruitment program. Certainly, insights from this study provide initial ideas for understanding these issues.

To assess of the effect of explanatory variables on the decision of willing to work as crowd-shippers or not, the marginal effects were calculated. Marginal effects other than elasticity were selected since the elasticity is generally used for measuring continuous explanatory variables and the majority of estimated variables in this research are indicator variables. In this study, the marginal effects measure the variation in the decision of working as a crowd-shipper as a function of a change in a certain variable, while keeping other variables constant. Of the total variables, the experience of transporting freight in the past and age greatly influence the WTW for crowdsource delivery companies. For example, experience with delivery freight increases WTW 18.2%, while all other variables remained the same. Moreover, the 30 years and older age group's WTW was 15% higher. The income variable has the least marginal influence on the WTW decision. An increase of $1,000 in annual individual income will lower the possibility of working as a crowd-shipper by 0.1%. All other variables have marginal effects in the range of 1-9%. All marginal effect coefficients are statistically significant, and have the same signs with the corresponding coefficients in the logit model.

**Tolerance of travel time model**

This section presents results from the corrected TTT regression model. The selectivity-bias approach is employed to correct for the TTT of respondents who were willing to work as crowd-shippers. Data from 1,720 observations (430 respondents) and discrete logit model outputs presented in Table 3 were employed to evaluate the regression model. Moreover, it is noteworthy to see the differences between the two models; therefore, results of the model estimated without the selectivity correction term are also presented in Table 4.

Regarding the model estimated with the selectivity correction term, the commuting trip parameter is negative and significant influence on the TTT. Respondents were willing to carry freight on their commuting trips but less likely to divert for longer times compared to other trip purposes. This finding is consistent with the fact that respondents may have more flexibility in their schedules in other contexts, e.g. during leisure trips or free time. Therefore, they can make a longer diversion to transport packages during these latter scenarios. On the other hand, the parameter of "expected to be paid as a crowd-shipper" is positive and significant. Thus, the more respondents are paid, the longer distances they are willing to travel. Considering this, the compensation schemes should be carefully designed to attract occasional drivers, but not to induce considerable vehicle miles traveled. Long extra driving by driver partners may overcompensate the resource savings (e.g. fuel consumption per package delivery); therefore, violate the objectives of implementing crowd-shipping systems that are improved mobility, safety, and environmental sustainability. One possible solution is to break-down long delivery trips so multiple crowd-shippers can cooperate to deliver the same request on their travel anyway. As such, crowd-shippers' route deviation is minimized.

During the design of the model, we were interested to identify potential crowd-shippers' package ownership preference. Interestingly, the coefficient of the variable for "I can deliver whoever's packages if I get paid" negatively influenced delivery TTT. As such, respondents are more likely to travel longer once they transport freight or goods for friends, colleagues, relatives, or neighbors. This suggests that crowd-shippers are more willing to divert from their routes to transport packages for people who are closely linked to them. One way to potentially improve the crowd-shipping market would be to link the crowd-shipping with individuals' social network. Similarly, young people (i.e. less than 31 years old) and females are willing to travel longer to deliver packages.

The results also clearly show that the African American, American Indian, Alaska native males parameter is positive and statistically significant. Therefore, this segment of the population are more likely to travel longer to deliver once they work as driver partners for crowd-shipping companies. Moreover, respondents with low incomes (i.e. less than $30,001/year) are likely to travel longer to deliver freight at night. This result suggests that low-income respondents are more likely to accept work at times that are unattractive to other people.

The parameters identified in both the models are worth noting. In the two models, all common parameters are found significant, except the "age" parameter. The "age" parameter is not significant in the model estimated without the selectivity correction term. Furthermore, in the random parameter model that is estimated with the selectivity correction term, the selectivity-bias parameters are statistically different from zero. As such, the selectivity correction parameter varies significantly across observations. Therefore, the null hypothesis of the selectivity-bias parameter equal to zero can be rejected at the confidence level of more than 99.99%. These results also concur with our sample selectivity hypothesis. Thus omitting the selectivity correction term leads to serious model misrepresentation. For instance, when comparing the two models estimated with and without the selectivity correction term, the parameters are remarkably different, especially the

**Table 4.** Corrected and un-corrected regression models of tolerance for travel time

| Variable Description | Estimate with selectivity correction | | Estimate without selectivity correction | |
|---|---|---|---|---|
| | Parameter | t-stat | Parameter | t-stat |
| **Non-random parameters** | | | | |
| Constant | 11.099 | 6.60 | 8.245 | 3.69 |
| I can be a crowd-shipper during my commute | -5.076 | -6.63 | -4.902 | -8.90 |
| ETP as a crowd-shipper | 4.988 | 23.22 | 5.043 | 25.37 |
| I can deliver whosoever packages or goods if I get paid | -3.976 | -4.84 | -4.374 | -4.48 |
| Age. Dummy variable: 1- if <31 years old; 0- otherwise | 2.322 | 2.37 | 0.340 | 0.37 |
| Female | 2.418 | 2.88 | 1.780 | 10.30 |
| African American/American Indian/Alaska native males | 8.564 | 4.64 | 10.216 | 4.06 |
| Having income is less than $30,000/year and willing to deliver at weekday nights | 1.991 | 2.23 | 2.370 | 2.66 |
| **Random parameters** | | | | |
| Mean of selectivity correction term | 5.936 | 4.39 | - | - |
| Standard derivation of selectivity correction term | 14.954 | 63.04 | - | - |
| Number of observation (respondents) | 1,720 (430) | | 1,720 (430) | |
| R square | 0.270 | | 0.261 | |
| Corrected R square | 0.266 | | 0.258 | |
| Number of Draws | 1000 | | 1000 | |
| **Computed values** | | | | |
| Expect to be paid (ETP) ($/h) | 12.029 | | 11.898 | |

*Note:* Insignificant parameters are underlined.

constant and "age" parameters. As such, when the selectivity bias terms are ignored, erroneous interpretation and conclusions are produced from the estimated results.

In this research, ETP is the amount crowd-shippers expect to be paid for their delivery driving time. This value is similar to the WTW value in Miller et al. (2017) (*15*). The ETP value of the model with selectivity correction is $12/hour, lower than the average WTW value reported by Miller et al. (2017) ($19/hour). However, this ETP value is within the $9.2 to $15.6 hourly value range of travel time saving published by the US Department of Transportation (*36*). The finding of an ETP value might suggest crowd-shipping companies to set compensation schemes that align with driver expectations. This will potentially increase the recruitment and retaining crowd-shippers in the system.

## CONCLUSIONS

Crowd-shipping or crowdsourced delivery companies provide platforms to connect senders who need to send packages to couriers who travel anyway. The system brings potential benefits to society, including improved mobility and reduced congestion and greenhouse gases. However, to implement an effective and efficient system, more understanding of the stakeholders, especially the crowd-shippers themselves, is needed. There is a lack of research on this topic; therefore, this paper addressed the central questions of identifying the factors that influence the behavior (WTW and TTT) of those interested in joining the crowd-shipping system. A survey has been conducted to collect data for the discrete-continuous model estimations. A binary logit model has been used to examine factors' influence on the WTW as crowd-shippers. An ordinary least-square regression model has been employed to understand the factors that affect the travel time decisions of potential crowd-shippers. The correlations of the discrete and continuous variables were corrected by a selectivity-bias term in the regression model. This correction is to prevent erroneous insights and conclusions derived from the results. Overall, the results show that the parameters have plausible signs and are statistically significant.

   The contributions of this research are of value to researchers, policy makers, and crowd-shipping companies. In summary, the contributions and suggested implementations are as follows:

- The use of discrete-continuous approaches that capture the maximum and random utility behaviors derived from heterogeneous samples. A selectivity-bias term included in the regression model corrects for the conditional selection behavior of potential driver partners' maximum TTT. Moreover, the statistical significance of the random selectivity-bias parameter confirmed the variation in respondent behavior.
- The findings for the main socio-demographic characteristics that influence prospective crowd-shippers' WTW may potentially help crowd-shipping companies to more successfully recruit employees. Future works should consider additional factors, such as package characteristics (e.g. weight and size), incentives, and scenario contextualization. As such, insights from the estimated results are helpful to assess the importance of variables and the circumstances in which individuals are willing to be driver partners. Those insights are also valuable for crowd-shipping companies' operational strategies (e.g. matching criteria). For example, the information help to match requests and couriers and potentially allow couriers to deliver goods around the clock and thereby avoid peak travel period.
- The use of incentives is a significant influence on the willingness of crowd-shippers to travel additional time for package pickup and delivery. ETP information is also helpful for crowd-shipping companies' operational strategies. For example, driver partner compensations can be designed based on the time of the day and the day of the week.
- It has potential of sharing the data (speed and travel time) collected by crowd-shipping firms, and integrating the data with daily transportation operation/management centers to improve the urban mobility, safety, and environment. By providing the data, crowd-shipping companies also build trust with regulators.
- Certainly, government bodies play a crucial role to grow a crowd-shipping industry through legislations, regulations, and subsidies. For example, provide appropriate incentive packages to attract ordinary drivers to switch to crowd-shipping driver partners are a feasible initiative.

   In conclusion, this research has provided for the first time important insights into the behaviors regarding the supply generation for crowd-shipping system. Future research is still needed to validate the findings in different contexts and extend the knowledge in this field.

## ACKNOWLEDGMENT

Authors are grateful for the suggestion of selecting the model from Professor Fred Mannering of the University of South Florida.